\documentclass[%
reprint,
superscriptaddress,
showpacs,preprintnumbers,
amsmath,amssymb,
aps,pre, 
nofootinbib,
longbibliography, 
floatfix,
]{revtex4-1}

\usepackage{enumitem}  
\usepackage{graphicx}
\usepackage{epstopdf}
\usepackage{dcolumn}
\usepackage{bm}
\usepackage{xcolor,soul}
\usepackage[none]{hyphenat}

\newcommand{\ie}{\textit{i.e.}\ }

\newcommand{\md}{\mathrm{d}}
\newcommand{\etal}{\textit{et al.}}

\begin{document}


\title{Theory of electromagnetic wave frequency upconversion in dynamic media}
\author{Kenan Qu}
\author{Qing Jia}
\affiliation{%
	Department of Astrophysical Sciences, Princeton University,  Princeton, New Jersey 08544, USA
}%
\author{Matthew R. Edwards}
\affiliation{%
	Department of Mechanical and Aerospace Engineering,
Princeton University, Princeton, New Jersey 08544, USA
}%
\author{Nathaniel J. Fisch}
\affiliation{%
	Department of Astrophysical Sciences, Princeton University,  Princeton, New Jersey 08544, USA
}%

\date{\today}

\begin{abstract}
	Frequency upconversion of an electromagnetic wave can occur in ionized plasma with decreasing electric permittivity and in split-ring resonator-structure metamaterials with decreasing magnetic permeability. We develop a general theory to describe the evolution of the wave frequency, amplitude, and energy density in homogeneous media with a temporally decreasing refractive index. We find that upconversion of the wave frequency is necessarily accompanied by partitioning of the wave energy into low-frequency modes, which sets an upper limit on the energy conversion efficiency. The efficiency limits are obtained for both varying permittivity and varying permeability. 
\end{abstract}

\maketitle

\section{introduction}
The ability to coherently convert the frequency of an electromagnetic wave between arbitrary wavelengths has a broad range of applications, including expanding the wavelengths accessible by current laser technologies. Frequency conversion is typically achieved using nonlinear optical methods, which often require high light intensities. As an alternative linear method, frequency conversion by temporally changing the medium refractive index~\cite{morgenthaler1958velocity, wilks1988frequency, mendoncca2000book, nerukh2012non, kalluri2016electromagnetics} has attracted particular attention for its ability to create ``tabletop'' sources of coherent x rays out of near-infrared lasers at high efficiency and low cost~\cite{Manuscript_Chirped, Dustin_np_2018}. A rapid change in refractive index has also been proposed~\cite{yablonovitch1989accelerating, 2013-PNAS} for observing the Unruh effect, which is crucial for testing fundamental quantum field theory. For these purposes, frequency upconversion has been experimentally demonstrated in the microwave~\cite{yablonovitch1973spectral, joshi1990demonstration, kuo1990frequency, savage1993frequency, yugami2002experimental}, terahertz~\cite{Nishida2012}, and optical~\cite{Suckewer2002, Suckewer2005, Shvets2017arxiv} regimes using dynamic plasmas and metamaterials~\cite{Shvets2017arxiv, 2013-PNAS}.

Despite the wide spectra and disparate materials that can realize frequency upconversion using dynamic media, previous theories have generally focused on plasma with an increasing density~\cite{wilks1988frequency, stepanov1993waves, bakunov2000adiabatic, kalluri2016electromagnetics}. These theories have neglected the electromagnetic wave interaction with atomic resonances before the medium is ionized, and thus overestimated energy conversion efficiency. Previous theories have also ignored metamaterials as dynamic media capable of inducing large frequency shifts using engineered refractive indices.

In this paper, we formulate a general theory describing frequency upconversion of electromagnetic waves in dynamic homogeneous media. Compared to previous theories, our study obtains general constitutive relations when polarization and magnetization suddenly change in the frequency conversion process.  
Our theory describes how the electromagnetic field changes its frequency, amplitude, and energy density when the medium changes its refractive index. We show that the process of frequency upconversion necessarily leads to the division of wave energy into multiple modes with components at both higher and lower frequencies than the initial frequency.

This article is organized as follows: In Sec.~\ref{sec:model}, we first introduce the theory of frequency upconversion of an electromagnetic wave starting from Maxwell's equations. Then, we find the continuous quantities during the medium change by considering both polarization and magnetization and obtain the field amplitudes after the wave frequency is shifted. The obtained results are applied to media with decreasing permittivity in Sec.~\ref{sec:E} to find the efficiency of energy conversion to the high-frequency mode. We specifically analyze ionized plasma media and supplement the known theory of energy conversion efficiency when the medium changes from gas or solid state to plasma. In Sec.~\ref{sec:B}, we study frequency conversion using magnetic metamaterials with split-ring resonator (SRR) structures~\cite{SRR} which have variable magnetic permeability. We find that both variable permittivity and variable permeability yield energy conversion efficiencies reaching $50\%$. In Sec.~\ref{sec:concl}, we summarize the key points of this paper.

\section{Model}\label{sec:model}

Consider an electromagnetic plane wave with electric field $\bm{E}$ and magnetic field $\bm{B}$ propagating in a homogeneous medium.
The field evolution is described by Maxwell's macroscopic equations, 
\begin{equation} \label{1}
\begin{aligned}
\nabla\cdot\bm{E} &= \frac{\rho}{\varepsilon_0},  
&\nabla\times\bm{E} &= -\frac{\partial}{\partial t}\bm{B},  \\
\nabla\cdot\bm{B} &= 0,  
&\nabla\times\bm{B} &= \mu_0\bm{J} + \varepsilon_0\mu_0 \frac{\partial}{\partial t}\bm{E}, 
\end{aligned}
\end{equation}
where $\varepsilon_0$ and $\mu_0$ are the permittivity and permeability, respectively, of vacuum and $\varepsilon_0\mu_0 = 1/c^2$. The 
total charge density $\rho$ and the total current density $\bm{J}$ include components 
\begin{align}
	\rho &= \rho_\mathrm{free} + \rho_\mathrm{pol}, \label {2} \\
	\bm{J} &= \bm{J}_\mathrm{free} + \bm{J}_\mathrm{pol} + \bm{J}_\mathrm{mag}, \label{3} 
\end{align}
where $\rho_\mathrm{pol} = -\nabla\cdot\bm{P}$ is the polarization charge density, $\bm{J}_\mathrm{pol}=\partial\bm{P}/\partial t$ is the polarization current, and $\bm{J}_\mathrm{mag} = \nabla\times\bm{M}$ is the magnetization current. Here, $\bm{P}$ is the polarization density and $\bm{M}$ is the magnetization density. Since the electric displacement is $\bm{D}=\varepsilon_0\bm{E} +\bm{P} = \varepsilon\bm{E}$ and the magnetic field is $\bm{H} = \bm{B}/\mu_0-\bm{M}  = \bm{B}/\mu$, the polarization and magnetization can be equivalently described by the permittivity $\varepsilon$ and the permeability $\mu$, respectively. The refractive index is defined as $n=\sqrt{\varepsilon\mu /(\varepsilon_0\mu_0)}$.

In our discussion, we focus on the polarization and magnetization effects of the media in response to an input electromagnetic wave. We neglect $\rho_\mathrm{free}$ and $\bm{J}_\mathrm{free}$ since they do not contribute to polarization or magnetization. We count the electrons in metals and plasmas as polarization charges because their collective oscillation in response to electromagnetic fields induces a polarization field according to the Drude model~\cite{Drude}. Then, we can write the wave equations as
\begin{align}
	\nabla^2 \bm{D} &= \frac{n^2}{c^2} \frac{\partial^2}{\partial t^2} \bm{D} + \varepsilon \dot{\mu} \frac{\partial}{\partial t} \bm{D}, \label{4} \\
	\nabla^2 \bm{B} &= \frac{n^2}{c^2} \frac{\partial^2}{\partial t^2} \bm{B} + \dot{\varepsilon} \mu \frac{\partial}{\partial t} \bm{B}, \label{5}
\end{align}
where $\dot{\varepsilon}$ and $\dot{\mu}$ denote time derivatives. 
The spatial operation and temporal operations in the wave equations (\ref{4}) and (\ref{5}) are separated. For homogeneous media, the right hand sides of Eqs.~(\ref{4}) and (\ref{5}) are spatially invariant, so the spatial operators can be reduced $\nabla^2\to -k^2$. When the medium changes in time, Eqs.~(\ref{4}) and (\ref{5}) describe damped harmonic oscillations of the fields $\bm{D}$ and $\bm{B}$ with a central frequency $\omega=ck/n$. The first-order time derivative terms of $\bm{D}$ and $\bm{B}$ describe damping, which causes a finite frequency spectrum when $\dot{\varepsilon}$ or $\dot{\mu}$ is nonzero. The spectral width decreases as $\dot{\varepsilon}$ and $\dot{\mu}$ approach zero. 
In the steady state $\dot{\varepsilon}=\dot{\mu}=0$, the wave equations (\ref{4}) and (\ref{5}) reveal a linear correspondence relation $\omega = ck/n$ between the wave frequency $\omega$ and the wave number $k$. Thus, an electromagnetic wave with a certain $k$ will change its frequency $\omega$ when the homogeneous refractive index $n$ changes in time. In the time scale that $n$ is defined, the wave frequency in the steady state is solely determined by $ck/n$ but does not depend on the history of $n$.

Assume an electromagnetic wave with initial frequency $\omega_\mathrm{i}$ propagates in a medium whose refractive index is dynamic. The final wave frequency is determined by the equation $\omega = ck/n$, which may have multiple solutions $\omega_j$ in dispersive media where $n(\omega)$ is frequency dependent. The initial wave energy can be projected into each of these frequency modes. To find out how the initial electric field amplitude $E_\mathrm{i}$ and magnetic field amplitude $B_\mathrm{i}$ are converted, we analyze the continuous quantities of the wave when the medium changes its refractive index, or specifically the polarization and magnetization.  

We first write the general steady-state solution for the fields as a superposition of multiple frequencies 
\begin{align}
	\bm{E} &= \sum_j (E_{j+} e^{-i\omega_j t} + E_{j-} e^{i\omega_j t}) e^{-ikz} \bm{e}_E, \label{6} \\
	\bm{B} &= \sum_j (B_{j+} e^{-i\omega_j t} + B_{j-} e^{i\omega_j t}) e^{-ikz} \bm{e}_B, \label{7} 
\end{align}
where  $\omega_j$ runs through all the solutions to $\omega=ck/n(\omega)$; $E_{j\pm}$ and $B_{j\pm}$ represent the amplitudes of the electric and magnetic fields at frequency $\omega_j$ in the forward- and backward-propagating directions, respectively. An electrostatic or magnetostatic mode can be described with $\omega_j=0$. It should be noted that the fields $\bm{E}$ and $\bm{B}$ oscillate in phase, and hence  the amplitudes $E_{j\pm}$ and $B_{j\pm}$ are expressed as real values. 

In the case of frequency upconversion, reducing the refractive index $n$, which is proportional to $\sqrt{\varepsilon\mu}$, is accomplished by increasing $\rho_\mathrm{pol}$, $\bm{J}_\mathrm{pol}$, and $\bm{J}_\mathrm{mag}$, which oscillate at the same frequency as the $\bm{E}$ and $\bm{B}$ fields. Examples of such processes include forming new molecules with larger polarizability or magnetizability, creating or increasing plasma density, and activating metamaterials that can interact with an electromagnetic wave. Wave frequency conversion arises from the coupling between the input electromagnetic wave and the oscillation of the new polarization and magnetization. 
For simplicity, we assume that the change in medium is triggered by an external field at $t=0$ which uniformly creates new charges $\Delta\rho$ and new currents $\Delta\bm{J}$ in a time shorter than an optical cycle $2\pi/\omega_\mathrm{i}$. The simultaneous creation ensures that the charges and currents begin to oscillate coherently under the drive of the input electromagnetic field, inducing polarization $\bm{P}$ and magnetization $\bm{M}$  and thereby reducing $\varepsilon$ and $\mu$. The initial values of the $\Delta\rho$ and $\Delta\bm{J}$ oscillations are both zero at $t=0$. We must emphasize that the thermal motion-induced $\Delta\rho$ and $\Delta\bm{J}$ are not correlated with the input electromagnetic wave, hence their average contribution to polarization or magnetization is zero. With $\Delta\rho$ and $\Delta\bm{J}$ both zero at $t=0$, the values of $\rho$ and $\bm{J}$ are continuous. 
Applying the continuity of $\rho$ and $\bm{J}$ to Gauss's law in Eqs.~(\ref{1}), we first find that the gradients of the $\bm{E}$ and $\bm{B}$ fields are continuous at $t=0$. For a plane wave in a homogeneous medium, the continuity of the gradients can be generalized to the $\bm{E}$ and $\bm{B}$ fields themselves.  Continuity of $\nabla\times\bm{E}$, $\nabla\times\bm{B}$ and $\bm{J}$ subsequently leads to the continuity of  $\partial\bm{B}/\partial t$ and $\partial\bm{E}/\partial t$ at $t=0$ using Faraday's law and Ampere's law in Eqs.~(\ref{1}).  Therefore, we obtain a set of continuity relations of the instantaneous fields,  
\begin{equation} \label{8}
 \left.\left\{ \bm{E}, \bm{B}, \frac{\partial\bm{E}}{\partial t}, \frac{\partial\bm{B}}{\partial t} \right\} \right|_{t=0^-} 
 = \left.\left\{ \bm{E}, \bm{B}, \frac{\partial\bm{E}}{\partial t}, \frac{\partial\bm{B}}{\partial t} \right\} \right|_{t=0^+}.
\end{equation}
Since $\bm{E}|_{t=0^-} = E_\mathrm{i} e^{-i\omega_\mathrm{i} t-ikz}$ and $\bm{B}|_{t=0^-} = B_\mathrm{i} e^{-i\omega_\mathrm{i} t-ikz}$, substituting $\bm{E}|_{t=0^\pm}$ and $\bm{B}|_{t=0^\pm}$, which obey the general solution (\ref{6}) and (\ref{7}) into the continuity relation (\ref{8}), yields 
\begin{align}
	E_\mathrm{i} &= \sum_j (E_{j+} + E_{j-}), \label{10} \\
	B_\mathrm{i} &= \sum_j (B_{j+} + B_{j-}), \label{11} \\
	\omega_\mathrm{i} E_\mathrm{i} &= \sum_j \omega_j(E_{j+} - E_{j-}), \label{12} \\
	\omega_\mathrm{i} B_\mathrm{i} &= \sum_j \omega_j(B_{j+} - B_{j-}). \label{13} 
\end{align}
We note that Eqs.~(\ref{11}) and (\ref{12}) can only be satisfied simultaneously when the output wave has multiple frequency components. Therefore, they set an upper limit on the efficiency of energy transfer into the component with the highest frequency. Calculating the exact energy conversion efficiency requires additional initial conditions since Eqs.~(\ref{10}) and (\ref{13}) are linearly dependent on each other.  

The second set of continuity relations can be found by analyzing the polarization and magnetization fields. At $t=0$, since $\rho$ and  $\bm{J}$ are continuous,  the instantaneous fields $\bm{P}$, $\bm{M}$, and $\partial\bm{P}/\partial t$ are continuous by definition. As a result, the fields $\bm{D}$ and $\bm{H}$ are also continuous.  However, $\partial\bm{M}/\partial t$, whose curl is proportional to $\partial \bm{J}_\mathrm{mag}/\partial t$, is obviously not continuous.  
To find the continuity relation of $\bm{M}$, we treat $\bm{J}_\mathrm{mag}$ as a number of charged particles in orbital motion. The magnetic momentum per atom or molecule is proportional to the product of particle charge and its angular momentum. The change in magnetic permeability can be a result of new charged particles joining the orbital motion or charged particles increasing their orbital momenta. In either scenario, the angular positions of the charged particles are continuous at $t=0$, \ie $\int_{-\infty}^0 \bm{M}(t') \md t'$ is continuous. Therefore, we obtain the second set of continuous quantities at the instant of medium change, 
\begin{multline} \label{9}
\left.\left\{ \bm{D}, \bm{H}, \frac{\partial\bm{P}}{\partial t}, \int_{-\infty}^0 \bm{M}(t') \md t' \right\} \right|_{t=0^-} \\
= \left.\left\{ \bm{D}, \bm{H}, \frac{\partial\bm{P}}{\partial t}, \int_{-\infty}^0 \bm{M}(t') \md t' \right\} \right|_{t=0^+}.
\end{multline}
By using $\bm{P} = (\varepsilon-\varepsilon_0)\bm{E}$ and $\bm{M} = (1/\mu_0 - 1/\mu)\bm{B}$, the continuity relation (\ref{9}) can be written as 
\begin{align}
	\varepsilon_\mathrm{i} E_\mathrm{i} &= \sum_j \varepsilon_j (E_{j+} + E_{j-}), \label{14} \\
	\frac{1}{\mu_\mathrm{i}} B_\mathrm{i} &= \sum_j \frac{1}{\mu_j}(B_{j+} + B_{j-}), \label{15}  \\
	\omega_\mathrm{i} (\varepsilon_\mathrm{i} - \varepsilon_0) E_\mathrm{i} &= \sum_j \omega_j(\varepsilon_j - \varepsilon_0)(E_{j+} - E_{j-}), \label{16} \\
	\frac{1}{\omega_\mathrm{i}} \bigg(\frac{1}{\mu_0} - \frac{1}{\mu_\mathrm{i}}\bigg) B_\mathrm{i} &= \sum_j \frac{1}{\omega_j} \bigg(\frac{1}{\mu_0} - \frac{1}{\mu_j}\bigg)(B_{j+} - B_{j-}), \label{17}  
\end{align}
where $\varepsilon_j$ and $\mu_j$  are the permittivity and permeability, respectively, of the medium at frequency $\omega_j$. 

Now, we have obtained eight continuity equations for the amplitudes of the electric and magnetic fields: Eqs.~(\ref{10})-(\ref{13}) and (\ref{14})-(\ref{17}). These equations are, however, not all independent. We first note that Eqs.~(\ref{10}) and (\ref{13}) are equivalent using relation $B_{j\pm} = \pm(k/\omega_j) E_{j\pm}$ considering $k/\omega_j$ is negative for the backward-propagating mode. Equation (\ref{16}) can be constructed from Eqs.~(\ref{12}) and (\ref{15}) by identifying $B_{j\pm}/\mu_j = \pm kE_{j\pm}/(\mu_j\omega_j)$. Thus, Eqs.~(\ref{13}) and (\ref{16}) can be removed from the initial conditions. 	In Eq.~(\ref{17}), the terms $B_\mathrm{i}/(\omega_\mathrm{i}\mu_\mathrm{i})$ and $B_{j\pm}/(\omega_j\mu_j)$ can be isolated and canceled by applying the identity $B_{j\pm}/(\omega_j\mu_j) = \varepsilon_j E_{j\pm}/k$ to Eq.~(\ref{14}), which simplifies Eq.~(\ref{17}) to 
\begin{equation}\label{18}
\frac{1}{\omega_\mathrm{i}} B_\mathrm{i} = \sum_j \frac{1}{\omega_j}(B_{j+} - B_{j-}). 
\end{equation}
Therefore, the electric and magnetic fields are determined by six independent relations: Eqs.~(\ref{10})-(\ref{12}), (\ref{14})-(\ref{15}) and (\ref{18}). Equations (\ref{10})-(\ref{11}) and (\ref{14})-(\ref{15}) explicitly describe the continuity of the fields $\bm{E}$, $\bm{B}$, $\bm{D}$, and $\bm{H}$. Equations (\ref{12}) and (\ref{18}), which describe the continuity of $\partial\bm{E}/\partial t$ and $\int_{-\infty}^0\bm{B}dt'$, are the consequences of the continuity of the incremental values $\rho_\mathrm{pol}$, $\bm{J}_\mathrm{pol}$, and $\bm{J}_\mathrm{mag}$. 


In media with dynamic $\varepsilon$ and $\mu$, the input electromagnetic field is coupled to the polarization and magnetization fields. They form a coupled three-mode system determining three eigenfrequencies with each containing two propagation directions. To find the wave amplitudes, we rewrite the continuity relations as six equations of $E_{j\pm}$ using the relations $B_\mathrm{i} = (k/\omega_\mathrm{i}) E_\mathrm{i}$ and $B_{j\pm} = \pm(k/\omega_j) E_{j\pm}$. Then the general solution for the wave amplitudes can be expressed in matrix form as 
\begin{multline} \label{20}
	\begin{pmatrix}
		E_{1\pm} \\ E_{2\pm} \\ E_{3\pm} 
	\end{pmatrix}
	 = \frac12 \left[
	 \begin{pmatrix}
	 	1 & 1 & 1 \\
	 	\varepsilon_1 & \varepsilon_2 & \varepsilon_3 \\
	 	1/\omega_1^2 & 1/\omega_2^2 & 1/\omega_3^2 
	 \end{pmatrix}^{-1}
	 \begin{pmatrix}
	 1 \\  \varepsilon_\mathrm{i} \\ 1/\omega_\mathrm{i}^2 
	 \end{pmatrix} \right. \\
	 \pm 	 \left.
	 \begin{pmatrix}
	 1 & 1 & 1 \\
	 1/\mu_1 & 1/\mu_2 & 1/\mu_3 \\
	 \omega_1^2 & \omega_2^2 & \omega_3^2 
	 \end{pmatrix}^{-1}
	 \begin{pmatrix}
	 \omega_1 \\  \omega_2/\mu_\mathrm{i} \\ \omega_3\omega_\mathrm{i}^2 
	 \end{pmatrix} \frac{1}{\omega_\mathrm{i}} \right] E_i,
\end{multline}
and $B_{j\pm} = \pm(k/\omega_j) E_{j\pm}$. 
If either $\varepsilon$ or $\mu$ remains constant when the medium changes, the coupled two-mode system determines four waves at two different eigenfrequencies whose amplitudes are given as 
\begin{align}
	E_{1\pm} &= \frac12 \bigg(\frac{\omega_\mathrm{i}^2 - \omega_2^2}{\omega_1^2 - \omega_2^2}\bigg) \frac{\omega_1}{\omega_\mathrm{i}} \bigg(\frac{\omega_1}{\omega_\mathrm{i}} \pm 1\bigg) E_\mathrm{i},  \label{21} \\
	B_{1\pm} &= \frac12 \bigg(\frac{\omega_\mathrm{i}^2 - \omega_2^2}{\omega_1^2 - \omega_2^2}\bigg) \bigg(1 \pm \frac{\omega_1}{\omega_\mathrm{i}}\bigg) B_\mathrm{i},  \label{22} \\
	E_{2\pm} &= -\frac12 \bigg(\frac{\omega_\mathrm{i}^2 - \omega_1^2}{\omega_1^2 - \omega_2^2}\bigg) \frac{\omega_2}{\omega_\mathrm{i}} \bigg(\frac{\omega_2}{\omega_\mathrm{i}} \pm 1\bigg) E_\mathrm{i}, \label{23} \\
	B_{2\pm} &= -\frac12 \bigg(\frac{\omega_\mathrm{i}^2 - \omega_1^2}{\omega_1^2 - \omega_2^2}\bigg) \bigg(1 \pm \frac{\omega_2}{\omega_\mathrm{i}}\bigg) B_\mathrm{i}. \label{24} 
\end{align}

The general solutions expressed in Eqs.~(\ref{20})-(\ref{24}) describe the amplitudes of the electromagnetic waves after the sudden change in the refractive index $n$ of a homogeneous medium. Since the polarization and magnetization respond to the input electromagnetic wave linearly, our results can be extended to media with slowly varying $n$ by dividing the change in $n$ into multiple intermediate steps. At each step, the wave frequencies $\omega_j$ are solely determined by the equation $\omega_j=ck/n(\omega_j)$, and the incremental changes in fields amplitudes $E_{j\pm}$ and $B_{j\pm}$ are related to their values at the previous step through Eqs.~(\ref{20})-(\ref{24}). Between the steps, the forward- and backward-propagating modes collect different phases $e^{\mp i\omega_j t}$. The phase difference between waves generated at different times causes constructive or destructive interference affecting the wave amplitudes. Thus, the field amplitudes $E_{j\pm}$ and $B_{j\pm}$ can be calculated by iterating the upconversion and propagation operations.    

With the wave amplitudes, one can first find that the total momentum $\bm{S}= \sum_j (E_{j+}H_{j+} - E_{j-}H_{j-})/c^2$ carried by the electromagnetic waves is generally not conserved. The result is not surprising because dividing a photon flow into multiple directions while maintaining its momentum would require an increase in total photon energy. We next find how wave energy is partitioned in each mode by specifying media with decreasing electric permittivity $\varepsilon$ and with decreasing magnetic permeability $\mu$. 


%

\section{Frequency upconversion with decreasing permittivity} \label{sec:E}

\begin{figure}[htp]
	\includegraphics[width=0.3\textwidth]{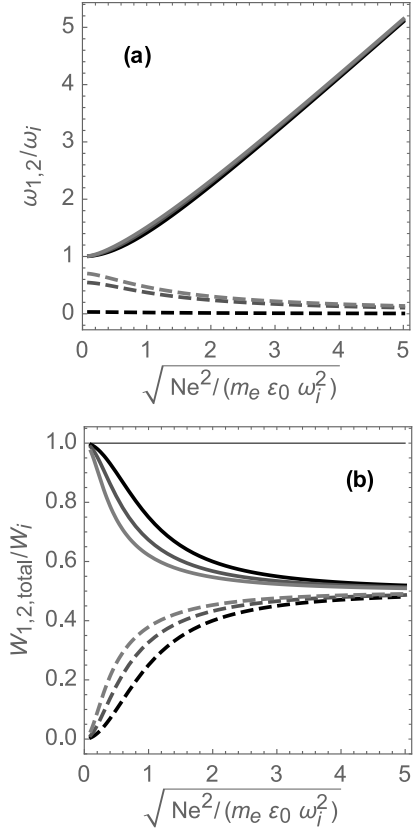}
	\caption{ (a) Frequency of each mode $\omega_1$ (solid) and $\omega_2$ (dashed) normalized to the initial wave frequency $\omega_\mathrm{i}$ and (b) energy density of each mode $W_1$ (solid), $W_2$ (dashed) and the total energy density $W_\mathrm{total}=W_1+W_2$ (thin black curve) normalized to the initial wave energy density $W_\mathrm{i}$ as a function of the coupling strength $\sqrt{Ne^2/(m_e\varepsilon_0)}$ normalized to $\omega_\mathrm{i}$. The three pairs of curves correspond to $\omega_0^2/\omega_\mathrm{i}^2= 0$ (black), $0.3$ (gray), and $0.5$ (light gray), respectively.  \label{figE}}
\end{figure}

We first consider an electromagnetic wave propagating in a homogeneous medium whose permittivity suddenly decreases. In general, permittivity is caused by the polarization of the atom and molecule or by currents in metals and plasmas. When polarization charges or current are suddenly created, they couple to the input electromagnetic field, increasing the wave frequency. Evolution of the $\bm{E}$ field and the $\bm{P}$ field can be described by
\begin{align}
	\nabla^2\bm{E} - \frac{1}{c^2} \frac{\partial^2\bm{E}}{\partial t^2} &= \frac{1}{c^2\varepsilon_0} \frac{\partial^2\bm{P}}{\partial t^2} \label{2.1} \\
	\frac{\partial^2\bm{P}}{\partial t^2} + \omega_0^2 \bm{P} &= \frac{Ne^2}{m_e} \bm{E}, \label{2.2}
\end{align}
where $\omega_0$ is the dipole resonance frequency ($\omega_0=0$ for metals and plasmas), $N$ is the number density of the dipoles, $e$ is the natural charge, and $m_e$ is the electron mass. Absorption can be neglected when in the course of frequency upconversion the wave frequency is kept away from $\omega_0$ by at least the atomic relaxation rate. With the coupling strength $\sqrt{Ne^2/(m_e\varepsilon_0)}$, the wave eigenfrequencies are found by Fourier transforming Eqs.~(\ref{2.1}) and (\ref{2.2}) to obtain
\begin{align}
	k^2\bm{E} - \frac{\omega^2}{c^2} \bm{E} &= \frac{\omega^2}{c^2\varepsilon_0} \bm{P} \label{2.1a} \\
	-\omega^2\bm{P} + \omega_0^2 \bm{P} &= \frac{Ne^2}{m_e} \bm{E}, \label{2.2a}
\end{align}
which on solving for $\omega$ yields 
\begin{multline} \label{2.3}
	\omega_{1,2} = \left[ \frac12 \bigg(c^2k^2 + \omega_0^2 + \frac{Ne^2}{m_e\varepsilon_0} \bigg) \vphantom{\sqrt{(\frac{Ne^2}{m_e\varepsilon_0})^2}} \right. \\
	 \pm \left. \frac12 \sqrt{\bigg(c^2k^2 + \omega_0^2 + \frac{Ne^2}{m_e\varepsilon_0}\bigg)^2 -4c^2k^2\omega_0^2} \right]^\frac12. 
\end{multline}
The same result can be obtained by solving the dispersion relation $k/\omega=\sqrt{\varepsilon(\omega)\mu}$ provided that $\varepsilon(\omega)$ is known. 
At a finite value of $N$, Eq.~(\ref{2.3}) reveals two eigenmodes: an electromagnetic wavelike mode with frequency $\omega_1$, and a dipole-oscillation-like mode with frequency $\omega_2$. The values of $\omega_{1,2}/(ck)$ are plotted in Fig.~\ref{figE}(a). 
The eigenfrequency $\omega_1$ increases at larger values of $N$. This correlation relation enables upconverting the frequency of a propagating electromagnetic wave by suddenly increasing the medium dipole density $N$. In the limit of large coupling strength $\sqrt{Ne^2/(m_e\varepsilon_0)}$, $\omega_1 \to \sqrt{c^2k^2 + \omega_0^2 + Ne^2/(m_e\varepsilon_0)}$ and $\omega_2 \to 0$. 

As the wave frequency $\omega_1$ increases and $\omega_2$ decreases, Eqs.~(\ref{21})-(\ref{24}) show that the electric fields $E_{1\pm} \to E_\mathrm{i}/2$ and $E_{2\pm} \to 0$ whereas the magnetic fields $B_{1\pm} \to \pm (\omega_\mathrm{i}/\omega_1) B_\mathrm{i}/2$ and $B_{2\pm} \to -B_\mathrm{i}/2$. The group velocity $v_{gj}$ can be found from Eq.~(\ref{2.3}),
\begin{equation} \label{2.7}
v_{gj} \equiv \frac{\partial\omega_j}{\partial k}= \frac{c^2k\omega_j Ne^2/(m_e\varepsilon_0)}{(\omega_j^2 - c^2k^2)^2 + c^2k^2 Ne^2/(m_e\varepsilon_0)}
\end{equation}
As the coupling strength increases, we find that the group velocity $v_{g1}$ decreases asymptotically to $c^2k/\omega_1$ whereas $v_{g2}$ approaches $\omega_2/k$. The medium impedance is $\eta_j = \sqrt{\mu/\varepsilon_j}$ which equals $\omega_j \eta_0/(ck)$ in non-magnetized media, where $\eta_0$ is the impedance of vacuum. Thus we can obtain the energy density in each mode, 
\begin{equation} \label{2.6}
W_j = \frac{1}{2v_{gj}\eta_j} (E_{j+}^2 + E_{j-}^2)
\end{equation}
In Fig.~\ref{figE}(b), we plot the energy density of each mode using Eq.~(\ref{2.6}) as well as the total energy density. From the figure, we observe that $W_1$ decreases monotonically whereas $W_2$ increases monotonically as the coupling strength $\sqrt{Ne^2/(m_e\varepsilon_0)}$ increases. The total energy density remains equal to $\varepsilon_0 E_\mathrm{i}^2$ which is the initial wave energy density $W_\mathrm{i}$ in vacuum. At large coupling strengths, both $W_1$ and $W_2$ approach $W_\mathrm{i}/2$ regardless of the frequency ratio $\omega_0/\omega_\mathrm{i}$.

The underlying physics can be explained by examining the coupling between the initial electromagnetic wave and the dipole oscillations. The dipole oscillation radiates in phase with the electromagnetic wave when $\omega_j<\omega_0$ and out of phase otherwise. Thus, destructive interference between the high-frequency mode $\omega_1$ and the dipole radiation causes a decrease in the wave amplitude. In the limit of large coupling strength, the wave energy is completely coupled into the dipole oscillation and is then radiated in both positive and negative directions. The radiation is also equally distributed in two modes with frequencies $\omega_1$ and $\omega_2$. Since the amplitude ratio of the electric field and the magnetic field is $\eta_j= |E_{j\pm}/H_{j\pm}|$, the energy in the higher- and lower-frequency modes is carried almost purely by electric fields and magnetic fields, respectively.

By setting the atomic resonance frequency $\omega_0$ to zero, our results can describe the reported phenomenon of electromagnetic wave frequency upconversion via suddenly ionized plasma~\cite{wilks1988frequency}. During ionization, the liberated electrons begin to oscillate at the plasma frequency $\omega_p = \sqrt{e^2n_e/(m_e\varepsilon_0)}$ leading to an increased wave frequency $\omega = \sqrt{c^2 k^2 + \omega_p^2}$. Here, $n_e$ denotes the number density of free electrons which can be larger than $N$ for fully ionized media. Note that plasma has no dipole frequency since the electrons are not bound to the atomic nucleus. The wave amplitudes can be found by reducing Eqs.~(\ref{21})-(\ref{24}) into     
\begin{align}
	\bm{E} &= (E_{1+} e^{-i\omega t} + E_{1-} e^{i\omega t}) e^{-ikz} \bm{e}_E, \label{4.51} \\
	\bm{B} &= (B_{1+} e^{-i\omega t} + B_{1-} e^{i\omega t} + B_2) e^{-ikz} \bm{e}_B, \label{4.52}
\end{align}
where
\begin{align}
	E_{1\pm} &= \frac12 \bigg(1 \pm \frac{\omega_\mathrm{i}}{\omega}\bigg) E_\mathrm{i},  \label{4.6} \\
	B_{1\pm} &= \frac12 \bigg(\frac{\omega_\mathrm{i}}{\omega} \pm 1\bigg)\frac{\omega_\mathrm{i}}{\omega} B_\mathrm{i}.  \label{4.7} \\
	B_2 &= \bigg(1- \frac{\omega_\mathrm{i}^2}{\omega^2}\bigg) B_\mathrm{i}. \label{4.9}
\end{align}
The field amplitudes are in agreement with Wilks \etal~\cite{wilks1988frequency}. The energy density of each mode can be found as
\begin{align}
	W_{1\pm} &= \varepsilon_0 (E_{1\pm})^2 = \frac14 \bigg(1+ \frac{\omega_\mathrm{i}^2}{\omega^2} \pm 2\frac{\omega_\mathrm{i}}{\omega}\bigg) \varepsilon_0 E_\mathrm{i}^2, \label{4.10} \\
	W_2 &=  \frac{\omega_p^2}{2 \omega^2} \varepsilon_0 E_\mathrm{i}^2. \label{4.11}
\end{align}
They include a forward propagating wave, a reflected wave, and a static magnetic field. Note that the energy density in each mode incorporates both the field energy and the electron mechanical energy. 
The energy partitioning in different modes can be verified by  the numerical simulation results reported in Ref.~\cite{Manuscript_Chirped}.  As the plasma frequency increases, the total energy density remains a constant $W_\mathrm{total} = W_{1+} + W_{1-}  + W_2 = \varepsilon_0 E_\mathrm{i}^2$. At high plasma frequencies, $W_{1+}$ and $W_{1-}$ approach $W_\mathrm{total}/4$, and $W_2$ approaches $W_\mathrm{total}/2$. They agree with the results in Sec.~3.4 of Ref.~\cite{kalluri2016electromagnetics}. 

However, our theory also reveals that the total energy density $W_\mathrm{total}$ equals the energy density of the initial wave only if the initial permittivity equals that of vacuum or plasma, \ie $\omega_0=0$. If the initial medium has a non-negligible $\omega_0$, the wave energy density is 
\begin{equation} \label{4.13}
	W_\mathrm{i} = \left[1 + \frac{Ne^2}{m_e \varepsilon_0} \frac{ \omega_0^2 }{(\omega_\mathrm{i}^2-\omega_0^2)^2} \right] \varepsilon_0 E_\mathrm{i}^2, 
\end{equation}
which is larger than the total energy density after the medium changes by the second term in the brackets. During frequency conversion, the energy stored in the dipole oscillation is lost from the electromagnetic wave.


In an experiment, one is often more interested in the propagation of a finite-duration pulse. When the pulse frequency is upconverted, its energy is partially reflected and lost. Since the amplitude of the reflection mode decreases proportionally to $(\omega_p/\omega_\mathrm{i})^2$ as $\omega_p$ increases, the reflection can be suppressed by gradually ionizing the plasma if the incremental plasma frequency in each step is much slower than $\omega_\mathrm{i}$, or $(d\omega_p/dt)/\omega_p \ll\omega_\mathrm{i}$. In this limit, the amplitude of the forward-propagation wave from step $s-1$ to step $s$ are related through Eq.~(\ref{4.6}) by $\displaystyle E_{1+}^{s} = \frac12\left(1+\frac{\omega^{s-1}}{\omega^s} \right) E_{1+}^{s-1}$ where $E_{1+}^s$ and $\omega^s$ are the wave amplitude and frequency at step $s$, respectively. By recursively applying the relation, we can obtain the output electric field in the asymptotic limit of gradually ionized plasma  
\begin{align}
	E_{1+} &= \prod_{s=1}^\infty \left[ \frac12 \left(1+\frac{\omega^{s-1}}{\omega^s} \right) \right] E_\mathrm{i} \nonumber \\
	&= \lim\limits_{\mathcal{N}\to\infty} \prod_{s=1}^\mathcal{N} \left\{ \frac12 \left[1+\frac{\omega_\mathrm{i} + (\omega-\omega_\mathrm{i})(s-1)/\mathcal{N}}{\omega_\mathrm{i} + (\omega-\omega_\mathrm{i})s/\mathcal{N}} \right] \right\} E_\mathrm{i} \nonumber \\
	&= \sqrt{\frac{\omega_\mathrm{i}}{\omega}}E_\mathrm{i}. 
\end{align}
Here, we use $\omega_\mathrm{i}$ and $\omega$ to denote the initial and final frequencies, respectively. 
In the last step, we used Euler's formula for the $\Gamma$ function $\Gamma(z) = \lim\limits_{\mathcal{N}\to\infty} \mathcal{N}! \mathcal{N}^z/\prod_{s=0}^{\mathcal{N}}(z+s)$. The energy density is then
\begin{equation}
W_{1+} = (\omega_\mathrm{i}/\omega) W_\mathrm{i}. 
\end{equation}
Therefore, the energy density of the wave decreases inversely with the wave frequency $\omega$. Although the same result in a gradually ionized plasma has been shown in previous studies~\cite{bakunov2000adiabatic, Ilya2010_1, Ilya2010_2} by treating the electromagnetic wave as the ponderomotive potential for the electron motion, our derivation has shown the connection between the sudden ionization limit and the adiabatic ionization limit.

\section{Frequency Upconversion with decreasing permeability} \label{sec:B}

Although the electric permittivity can be controlled through coupling the wave to the atomic or molecular polarization, tunable magnetic permeability has been realized prevalently in artificial materials. In this section, we investigate the use of SRR structures~\cite{SRR} for upconverting the frequency of an electromagnetic wave. We consider a specific type of SRR whose permeability is controlled by an external  field. For simplicity, we neglect the change in permittivity. 

Consider a microwave pulse propagating in a metamaterial with uniformly distributed SRR structures oriented perpendicular to the magnetic field of the microwave. We assume that the SRRs begin to interact with the microwave pulse only when activated by an external field at time $t=0$. Once activated, a loop current $I$ is induced in the rings when the magnetic flux changes. For each ring, the split gap stops the current and helps accumulate positive and negative charges on two sides of the gap, thus forming a capacitor with capacitance $C$. Variation of the loop current at the microwave frequency is also subject to self-inductance $L$ and a mutual inductance $-FL$. Here, $F$ is the area filling fraction and denotes the coupling strength between the electromagnetic wave and the SRR structure. 
Thus, the SRR can be modeled as an LC circuit and the current $I$ satisfies 
\begin{equation} \label{5.1}
(1-F)L \frac{\md I}{\md t} + \frac1C \int^t I\md t' = -\frac{\md}{\md t}(F a^2 B),
\end{equation}
where $a$ is the lattice constant of the SRR structure. 
Using relation $LI=\mu_0 M a^2$, it becomes 
\begin{equation} \label{5.2}
\frac{\partial^2}{\partial t^2} M + \omega_0^2 M = -F \frac{\partial^2}{\partial t^2}(\frac{1}{\mu_0}B-M),
\end{equation}
where $\omega_0 = 1/\sqrt{LC}$ is the resonance frequency of the LC circuit. 
Without $B$, the effective resonance frequency of the magnetization $M$ is $\omega_0/\sqrt{1-F}$ which increases when $F$ approaches $1$. 
During the change in medium, the continuity relations are given by the same equations (\ref{8}) and (\ref{9}). Here, the continuity of $\int_{-\infty}^0 \bm{M}(t')\md t'$ is caused by the fact that the SRRs are not charged initially, \ie $\int_{-\infty}^0 I\md t'=0$. 

\begin{figure}[htp]
	\includegraphics[width=0.3\textwidth]{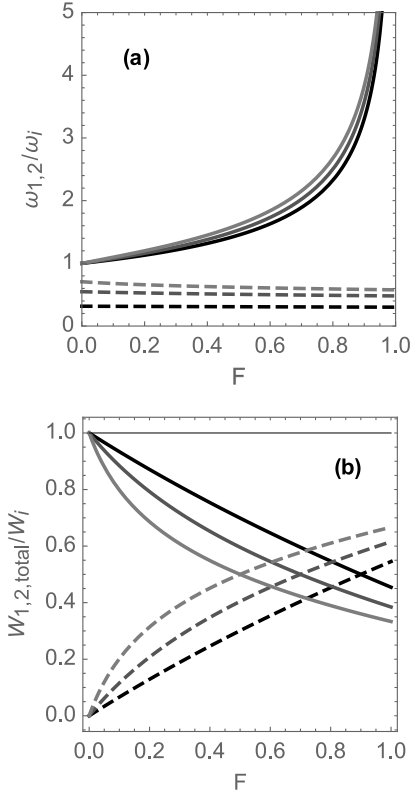}
	\caption{ (a) Frequencies of each mode  $\omega_1$ (solid) and $\omega_2$ (dashed) normalized to the initial wave frequency $\omega_\mathrm{i}$ and (b) energy density of each mode $W_1$ (solid), $W_2$ (dashed) and the total energy density $W_\mathrm{total}=W_1+W_2$ (thin black curve) normalized to the initial wave energy density $W_\mathrm{i}$ as a function of the filling fraction $F$. The three pairs of curves correspond to $\omega_0^2/\omega_\mathrm{i}^2= 0.1$ (black), $0.3$ (gray), and $0.5$ (light gray), respectively.  \label{figB}}
\end{figure}

The wave equation for the magnetic field can be obtained from Maxwell's equations (\ref{1}) using $\bm{J}=\nabla\times\bm{M}$  
\begin{equation} \label{5.3}
\nabla^2 B - \frac{1}{c^2}\frac{\partial^2}{\partial t^2} B = \mu_0\nabla^2 M.
\end{equation}
The frequency of an electromagnetic wave in an SRR structure can be analyzed by taking the Fourier transform of Eqs.~(\ref{5.2}) and (\ref{5.3}) to find
\begin{multline} \label{5.4}
\omega_{1,2} = \bigg\{ \frac{1}{2(1-F)} \bigg[ (c^2k^2 + \omega_0^2)  \\
	 \pm \sqrt{(c^2k^2+\omega_0^2)^2 - 4(1-F)c^2k^2\omega_0^2} \bigg] \bigg\}^{1/2}. 
\end{multline}
The same result can be obtained by solving the dispersion relation $k/\omega=\sqrt{\varepsilon\mu(\omega)}$ where $\mu=1-F\omega^2/(\omega^2-\omega_0^2)$ in such materials~\cite{SRR}. 
The frequencies $\omega_{1,2}$ as a function of filling fraction $F$ is plotted in Fig.~\ref{figB}(a). 
It is evident that an increased coupling strength $F$ with any finite SRR resonance frequency $\omega_0$ leads to an upshifted eigenfrequency $\omega_1$ and a downshifted $\omega_2$. In the limit $F \to 1$, they approach, respectively,
\begin{align}
	\omega_1 &\to \sqrt{(c^2k^2 + \omega_0^2)/(1-F)}, \label{5.5} \\
	\omega_2 &\to \sqrt{c^2k^2\omega_0^2/(c^2k^2 + \omega_0^2)}. \label{5.6} 
\end{align}
Compared with Eq.~(\ref{2.3}) where $\omega_1^2$ increases linearly with the coupling strength when $\varepsilon$ is varied, we find that $\omega_1^2$ in Eq.~(\ref{5.4}) scales inversely with $1-F$ when $\mu$ is varied.  The special scaling relation of $\omega_1^2$ arises from the existence of mutual inductance $-FL$ of SRRs which combines with the self inductance $L$ causing a near-zero effective inductance $(1-F)L$ when $F\to1$. Then it elevates the effective SRR resonance frequency as well as the eigenfrequency $\omega_1$. 

In a medium with varying $\mu$ but constant $\varepsilon$, the material impedance becomes $\eta_j = ck\eta_0/\omega_j$. The group velocity of each mode is 
\begin{equation}
v_{gj} = \frac{\omega_1^2\omega_2^2}{2\omega_j^2-(\omega_1^2+\omega_2^2)} \left(\frac{1}{\omega_0^2}- \frac{1}{\omega_j^2} \right) \frac{\omega_j}{k},
\end{equation}
for $j=1,2$. 
Assuming $\omega_\mathrm{i}=ck$, the energy densities of each mode with different values of filling fraction $F$ and $\omega_0$ are plotted using Eq.~(\ref{2.6}) in Fig.~\ref{figB}(b). We observe that an increasing value of $F$ causes a decrease of energy density of the high-frequency mode $W_1$ and an increase in the energy density of the low-frequency mode $W_2$. The total energy density $W_\mathrm{total}$ is conserved.

As $\omega_1$ increases and $\omega_2$ decreases, the high-frequency mode becomes purely electric, \ie $E_{1\pm} \to (\omega_2^2/\omega_\mathrm{i}^2 -1)E_\mathrm{i}$ and $B_{1\pm} \to 0$; The low-frequency mode remains electromagnetic and $B_{2\pm} \to (1 \pm \omega_2/\omega_\mathrm{i}) B_\mathrm{i}/2$. The wave energy partially resides in the current oscillations of the split rings. 
At the maximum coupling strength with $F\to 1$, the energy density of each mode approaches 
\begin{equation}
W_{1,2} \to \frac12 \left(1\mp \frac{\omega_0^2}{\omega_\mathrm{i}^2+\omega_0^2} \right) W_\mathrm{i}, 
\end{equation}
where $W_\mathrm{i} = \varepsilon_0 E_\mathrm{i}^2$ is the initial wave energy density in vacuum. 
The result shows that the efficiency of energy conversion into the high-frequency mode is generally lower than $50\%$ for a finite value of $\omega_\mathrm{i}/\omega_0$ although the efficiency asymptotically approaches $50\%$ as $\omega_\mathrm{i}/\omega_0$ increases.



For frequency upconversion with simultaneously decreasing permittivity and permeability, it is straightforward to calculate the generated mode frequencies by solving the dispersion relation $k/\omega=\sqrt{\varepsilon(\omega)\mu(\omega)}$. We do not provide the explicit form of the energy density here as it is lengthy but not illuminative. We can expect a large frequency upshift resulting from the coupling between the upconverted oscillation frequencies of the electric field and the magnetic field. The energy density would be divided into three modes: an electromagnetic mode at upconverted frequency, a dipole-like mode at a low frequency, and a ring-shaped oscillating current mode at a low frequency.

\section{conclusion} \label{sec:concl}

To conclude, we have shown that upconverting the frequency of an electromagnetic wave using a temporally changing refractive index is accompanied by energy loss to low-frequency modes. We analytically obtained the field amplitudes and energy densities after frequency upconversion in a homogeneous medium when an external field causes sudden refractive index reduction. We have also outlined how our theory can be extended to media with gradually changing refractive index and how to calculate the modes amplitudes by iterating the sudden change processes. Our results show that media with decreasing permittivity $\varepsilon$ and with decreasing permeability $\mu$ have comparable energy conversion efficiencies, but the upconverted frequencies scale differently with the coupling strengths. The difference is rooted in the special structure of SRRs that a larger filling fraction $F$ not only increases the coupling strength between the magnetic field and the magnetization current, but also elevates the effective resonance frequency of SRRs.

Experimental observations of frequency upconversion using dynamic media are subject to technological constraints. Using ionized plasma, the most prominent difficulty is to create high density plasma in few optical cycles. For this reason, the most successful demonstration achieved only about $3.5-\mathrm{THz}$ upconversion by ionizing semiconductors~\cite{Nishida2012} using a moderate intensity ($4\times 10^9-\mathrm{W/cm}^2$) laser. Frequency upconversion in the optical range needs higher plasma density, which might be achieved using high intensity lasers. Using metamaterials, the foreseeable obstacle lies at the low resonance frequency limited by the fabrication technology. Even with the state-of-the-art technology, structures at the precision of micrometers~\cite{ssr_review} can only be fabricated, which limits the resonance frequency to the range of terahertz. Frequency upconversion beyond the optical frequency will require either advances in the nanofabrication technology or using a cascaded sequence of sudden refractive index change.

\begin{acknowledgments}
	This work was supported by NNSA Grant No. DE-NA0002948, and AFOSR Grant No. FA9550-15-1-0391.
\end{acknowledgments}  

\bibliography{upshift}

\end{document}